# WEB ACCESS TO CULTURAL HERITAGE FOR THE DISABLED

**Jonathan P. Bowen**
Professor of Computing, FEST/CISM
London South Bank University
Borough Road, London SE1 0AA, UK
E-Mail: jonathan.bowen@sbu.ac.uk   &   URL: http://www.jpbowen.com/

**Abstract** – Physical disabled access is something that most cultural institutions such as museums consider very seriously. Indeed, there are normally legal requirements to do so. However, online disabled access is still a relatively novel and developing field. Many cultural organizations have not yet considered the issues in depth and web developers are not necessarily experts either. The interface for websites is normally tested with major browsers, but not with specialist software like text to audio converters for the blind or against the relevant accessibility and validation standards. We consider the current state of the art in this area, especially with respect to aspects of particular importance to the access of cultural heritage.

> *O world invisible, we view thee,*
> *O world intangible, we touch thee,*
> *O world unknowable, we know thee,*
> *Inapprehensible, we clutch thee!*
>
> — Francis Thompson (1859–1907)

## INTRODUCTION

A cartoon in *The New Yorker* magazine on 5 July 1993 showed a picture of a dog at a computer talking to his friend and saying:

> *On the Internet, nobody knows you're a dog.*
> – Peter Steiner

A significant number of people using the Internet in general and the web in particular have some form of disability that may affect their use of the technology. Of course the World Wide Web Consortium (W3C), the main web standards body, is aware of the issues [15] but many sectors still have little awareness of the access problems to their websites. Museums and related cultural institutions normally pride themselves on the accessibility of their physical buildings with expensive lifts, induction loops for audio access, special toilet facilities, etc. – see in Figure 1 for example, the prominent ramp for the River and Rowing Museum at Henley that is raised well above ground level which is in the flood plain of the River Thames. It would not be untypical to spend around 10% or more of a building's overall cost on improved accessibility. This is both a legal requirement in many countries and also a moral duty for public-spirited institutions such as museums. What is more, perhaps around 10% of the population have some sort of physical disability that impairs their activities in some way. However, many heritage institutions have yet to make an equivalent effort for their online facilities, despite the fact that legislation covering this mode of access is in the offing or already exists in most developed countries.

**Figure 1.** Disabled access ramp for a museum.

Unfortunately many web designers come from a graphic design background and thus may not be expert in non-visual forms of access. In addition, web tools do not necessarily enforce or even greatly aid improved accessibility. Indeed some are positively detrimental because their output often does not meet the basic HTML coding standards laid down by the World Wide Web Consortium (W3C). Instead they tend to aim at the latest web browsers and try to maximize the use of new features, with little concern for backward, cross-browser or cross-platform compatibility issues, let alone the problems of accessibility. To confound the issue, customers, not unreasonably, know ever less about the issues that the designers. What is more, the situation is getting worse because of the increasingly diverse set of technologies available on the web and rampant "featurism" as commercialisation increases. Of course, this is somewhat overstating the case and there are pockets of highly accessible web material from both museums and other organizations. Awareness is improving and most cultural heritage institutions are keen to improve the accessibility of their web facilities once they understand the issues and the fact that it is possible to do something about it, often at little cost if it is considered during a major website redesign.

2003 is the European Year of people with disabilities [www.eypd2003.org] and with the increasing importance of the web as a communication medium, cultural institutions should ensure the widest possible accessibility of their resources online. This does not *just* mean making cultural information available online in any form, as is thought by some less informed designers. It means thinking about the issues of different means of access (whether by blind, partially sighted, deaf, paralysed or otherwise disabled users).

As multimedia access increases and is improved through higher-speed access to the Internet and faster computers with better facilities, it is important that several modes of access are available for different users. For example, if audio is provided, a text transcript should also be available. If an image or video is presented, a text description and perhaps subtitles should accompany it.

Accessibility is a special case of *usability*. Computer users have understood that the "friendliness" of computers could be much improved for decades [16] and of course the technology has progressed remarkably with windows-based display, interactive keyboard, mouse access, stereo sound, etc., now being the norm. Improved usability aims to make use of technology (such as the web) more efficient, enjoyable, easier to remember, etc. Improved accessibility explicitly aims at widening the number of people who can use the technology, minimising the barriers that will always exist, but that can also always be reduced. Jakob Nielson provides an excellent website on web usability issues in general [www.useit.com], as well as writing leading books on the subject (e.g., see [10]). Here we concentrate on accessibility, in particular to cultural material where commercial pressures may be less and where there is a moral as well as legal imperative to widen accessibility.

## LEGAL ISSUES

At some point, it is likely that having an accessible website will become an accepted legal requirement for public institutions in most countries. As an example, the original website for the 2000 Olympics in Australia was not accessible for the disabled, especially the blind. A complaint was made in the case of Bruce Maguire, a blind person, versus the Sydney Organizing Committee for the Olympic Games (SOCOG)

[www.contenu.nu/socog.html], in which a single individual won against a large organization under the Australian Disability Discrimination Act. The case, known as "Maguire vs. SOCOG", was the first of its kind in the area of web accessibility. The statement of Tom Worthington, one of two expert witnesses, is available online [www.tomw.net.au/2000/mvs.html] together with a paper [17]. The Olympic website was actually deemed unlawful and the Australian Human Rights and Equal Opportunity Commission (HREOC) [www.hreoc.gov.au] ordered that it be made accessible to the disabled in time for the Olympics. However SOCOG ignored the ruling because its partner IBM said it would be too expensive and time-consuming to update the site, and thus were subsequently fined $20,000 (Australian), around £8,000, which is a fairly small sum for such an organization. However the effort involved in correcting the site was also in dispute. An irony is that the International Paralympic Committee, in parallel with the International Olympic Committee, even organizes games explicitly for the disabled [www.paralympic.org]. On the positive side, it is intended that the accessibility of the website for the 2004 Olympics games in Athens [www.athens2004.com] will be much improved as a result. However a check with the Bobby accessibility checker (see later) revealed three "Priority 1" (top priority) errors with missing alternative text for three images in the main homepage, a very basic accessibility mistake that is easy to correct.

In the United Kingdom, the Disability Discrimination Act (DDA) of 1995 [www.hmso.gov.uk/acts/acts1995/1995050.htm] applies to many bodies including museums and other cultural institutions. The UK government's disability website gives a useful and more friendly introduction to the Act [www.disability.gov.uk/dda]. This phased legislation will come fully into force in October 2004. Essentially, disabled access to services should be provided by institutions where possible. Part III of the Act (covering "Goods, facilities and services" amongst other aspects) makes it illegal for a service provider to treat those that are disabled in a less favoured manner because of their disability. "Information services" are explicitly covered under section 19(3) of the Act. Services should be adjusted by reasonable means to ensure that they are not impossible or unduly difficult to access, as covered by section 21(1) on the duty of providers:

> ***21. - (1) Where a provider of services has a practice, policy or procedure which makes it impossible or unreasonably difficult for disabled persons to make use of a service which he provides, or is prepared to provide, to other members of the public, it is his duty to take such steps as it is reasonable, in all the circumstances of the case, for him to have to take in order to change that practice, policy or procedure so that it no longer has that effect.***

An associated Code of Practice [6] is also available from the Disability Rights Commission (DRC) [www.drc.gov.uk]. This is a more approachable document than the Act itself and may be a better place to start for a non-lawyer needing some general guidance on the legal aspects of service provision.

More recent legislation explicitly covers web services under the Special Education Needs and Disability Act 2001 (SENDA), available from HMSO online [www.hmso.gov.uk/acts/acts2001/20010010.htm]. The new rights came into force on 1 September 2002, with some exceptions, including the provision of auxiliary aids and services that will be covered from 1 September 2003. This Act, essentially an extension of the DDA as mentioned above, is particularly aimed at protecting disabled students in higher and further education within the UK, so university museums and other institutions providing material for learning support will need to be compliant. An overview of what the Act will mean in practice, in a far more readable form than the Act itself, is available from the UK Centre for Legal Education (UKCLE) [www.ukcle.ac.uk/directions/issue4/senda.html].

There has been no court case involving website accessibility in the UK to date and it is unlikely that a museum would be sued, but it is obviously still incumbent on cultural institutions to act responsibly with respect to their website provision. The best point to act is the next time a major website redesign is planned, in which case the designers should be well aware of the legal issues of accessibility in the country concerned, as well as the technical and other solutions that are available to tackle the problems. For further information on and links to UK legislation, see [12]. For more general information on legislation internationally, see Chapter 2 of [14]; for US law, see Chapter13 in the same book and Appendix C for the US Section 508 Guidelines on accessibility of electronic and information technology that apply to US Federal agencies. See also Appendix A of [5].

## EXAMPLE MUSEUM WEBSITES

It is instructive to consider some museum sites where accessibility has been considered. This is now an aspect that is assessed in the Museums and the Web conference Best of the Web awards [www.archimuse.com/mw2003/best]. As an example, the Natural History Museum of Los Angeles County in the US [www.nhm.org] was initially an exemplary website from the point of view of accessibility. This was largely because someone who is very knowledgeable of the issues involved designed and organized it in-house over a number of years. Perhaps the most interesting thing about this and other websites designed with accessibility in mind is that they need look no different on a modern graphical web browser from any other professionally designed website. This demonstrates that designing with accessibility in mind does not mean one has to compromise what is on offer for the able bodied user with good web browsing facilities. Unfortunately the accessibility expert who used to be at this particular museum has since moved so the newly designed website is a retrograde step in regard to accessibility.

The British Museum COMPASS database, presenting a selection of the museum's best objects [www.thebritishmuseum.ac.uk/compass], includes a prominent "TEXT ONLY" link at the top of its main page for disabled users to gain easy access to the available facilities. The information for the graphical and text-based information is served from the same database, thus ensuring that both are in step and up to date. The British Museum is also developing a general text only are for its website, although this is still rather rudimentary [www.thebritishmuseum.ac.uk/text_only].

The Tate Gallery in London initiated the i-Map Project [www.tate.org.uk/imap] in associated with a major exhibition on Matisse and Picasso in 2002 [7]. This was designed to give access for visually impaired people via the web using raised images allowing them to be touched if printed on a special printer. Thus even art galleries that are obviously very visually oriented in general can make efforts to reach out to the blind.

The Imperial War Museum, also in London, has a text only version of the "Citizenship" area of its website under the auspices of its Education Services department [www.iwm.org.uk/education/citizenship]. There is also a text only page giving information for disabled visitors [www.iwm.org.uk/lambeth/disainfotxt.htm]. However, finding these from the main homepage is difficult, if not impossible. It is always worth making it obvious that accessible material is available in a clearly labelled manner, both to make it easier for the disabled online visitor, and to demonstrate that the effort to make the material available has been made.

More generally, the UK Heritage Lottery Fund provide an exemplary text only version of the website, linked from the homepage [www.hlf.org.uk]. This seems to be quite extensive, fast loading and easy to use, with a link back to the graphical version at the bottom of each page. The Virtual Library museums pages (VLmp) include an initial graphical hyperlink – with appropriate "alternative" text for text readers – which links to a text-based version of the complete website, dynamically generated using the Betsie tool (see later) [icom.museum/vlmp].

The Disability History Museum [www.disabilitymuseum.org], a completely virtual "museum", includes a "Text Only" link and is approved by the Bobby tool that checks for website accessibility (see next section). Another example of a real museum with a "text only" link is the Florida Museum of Natural History, USA [www.flmnh.ufl.edu]. However, further links return to standard graphical pages, which is not ideal. The Neuberger Museum of Art at the State University of New York, USA, makes a better attempt with a whole set of text only web pages available.

A museum with a good text version homepage, linked from the top of the main homepage, is the Smithsonian National Air and Space Museum in Washington DC [www.nasm.si.edu]. However, again the links from this page lead to graphical pages on the rest of the site.

Positioning the text link as the first link on the homepage is the ideal position for it to be found by blind visitors who must scan web pages sequentially using a text to audio conversion program such as JAWS (Job Access With Speech) for Windows from Freedom Scientific [www.freedomscientific.com]. Note that the font size for the text only link can be discreetly small if desired since the size does not affect the way the text is read. JAWS is available for evaluation free of charge (in a version that times out 30 minutes after each reboot of the computer on which it is run). This is perfectly adequate for demonstration purposes. It is highly recommended that web designers and museum personnel listen to their web pages being read by such software to gain an idea of the difficulties encountered by blind users.

The Rural History Centre (including the associated Museum of English Rural Life) at the University of Reading is accessible in text only form via the Betsie tool [access.museophile.net/www.ruralhistory.org]. This software was originally developed by the BBC, who are exemplary in their website accessibility, despite having a graphically rich site [www.bbc.co.uk].

Betsie is now "open source" software and is freely available to any organization that wishes to install it [betsie.sourceforge.net], although it requires a little expertise to do so. The software displays existing web pages by filtering their content to a version only displaying text in a uniform size, font and colour (see Figure 2). Betsie allows users to select text/background colour combinations, the font size and style on a special web page accessible via a link from the bottom of all web pages generated by the tool (see Figure 3). Thus it may be useful to partially sighted people requiring large size text as well as colour-blind people and those that are completely blind. A further link is provided so the user can return to the standard version of pages displayed by Betsie at any time. An advantage of Betsie is that the text only version of the website is dynamically generated from the normal web pages so there is no problem of ensuring that text-based pages are up to date. It can be very annoying to disabled users to be presented with inferior outdated information; so much so that they may actually prefer to attempt to access the main site anyway, even if it is less accessible.

**Figure 2.** Example web page generated by the Betsie tool.

**Figure 3.** Betsie options page.

## SURVEY AND TOOLS

The Bobby validator [bobby.watchfire.com], which can check for WAI compliance [15] and also for US Government Section 508 compliance [www.section508.gov], has been used to evaluate the accessibility and usability of 25 UK and 25 international museum and related websites with respect to their disabled accessibility (and hence usability) [3][4][8][9]. Bobby evaluates web pages for accessibility to users with disabilities. It checks for the presence or absence of particular features, or their characteristics, although it does not explicitly check HTML syntax. The W3C validator is recommended for checking HTML itself [validator.w3.org]. The results of this survey can be found in [9].

As well as the mechanical check using Bobby, a visual analysis was carried out manually using information provided by Bobby and also with textual browsing (e.g., using an audio browser) and partial sightedness in mind. Generally the sites faired quite badly with a significant number exhibiting some serious accessibility problems. The overall results for Bobby are shown in Table 1, extracted from [9]. Priority 1 errors are most serious and must be corrected to meet the WAI guidelines [15]. Priority 2 errors should be corrected if possible and Priority 3 errors may be corrected.

**Table 1.** Bobby validation results for museum websites.

The Bobby tool is available for free use on individual web pages via an online web interface (at a maximum rate of one page per minute to avoid misuse). A downloadable tool is also available at a charge for all major operating systems. An alternative, free, but more experimental web accessibility verifier (for use under Windows) is available for download from the A-Prompt ("Accessibility Prompt") Project at the University of Toronto, Canada [www.aprompt.ca].

A useful tool for obtaining a text view of any web pages is the Lynx browser [lynx.browser.org], available for Windows and Unix. A Lynx viewer website exists online, which saves having to load the Lynx software on a local computer [www.delorie.com/web/lynxview.html]. Any web page that is viewable satisfactorily using the Lynx browser is likely to be viewable by any web browser, so it is a useful check on the accessibility of web pages in practice. Just adhering to the accessibility standards is not enough to check all possible problems. As well as checks using Lynx and JAWS, for example, it is also highly recommended to try to have a disabled person access any new website, performing set tasks as well is free usage, ideally with the web designers present, and too give comments during and after the exercise.

## OTHER INFORMATION SOURCES

An excellent guide with museums specifically in mind is the Ed-Resources.Net Universal Access website by Jim Angus [www.ed-resources.net/universalaccess]. This includes an interesting comparison of the accessibility of three museum websites, as well as links to online web page validation services and further relevant resources.

In the UK, MAGDA, the Museums & Galleries Disability Association [www.magda.org.uk] is dedicated to improving access to UK museums and galleries in general for people with disabilities, as well as disseminating current best practice. It also provides an online forum for museum and gallery professionals to enable online discussion [groups.yahoo.com/group/magdamail].

The Jodi Mattes Access Award has been launched in 2003 to encourage improved accessibility to museum websites by offering formal recognition to the most accessible site as judged by a panel of representatives from relevant institutions [www.museumscomputergroup.org.uk/index_award.htm]. The award is endorsed by the UK Royal National Institute for the Blind (RNIB) [www.rnib.org.uk] as well as the UK Museums Computer Group (MCG) and MAGDA. It is named after Jodi Mattes (1973–2001) who was instrumental in the improved accessibility of the British Museum COMPASS resource (see earlier) amongst other projects. Hopefully this award will help to raise the awareness of online accessibility issues further in the cultural field by providing an incentive in the form of peer-reviewed recognition. The RNIB provide particularly good accessibility information on their website [www.rnib.org.uk/access], including good website design [www.rnib.org.uk/digital].

The number of books explicitly covering web accessibility has been very limited until recently, but at least four are now available. [11] was the first book in the area known to the author. [14] is written by a range of experts, including legal aspects in some detail for example, and as such is perhaps the most authoritative book in this area to date. [5], the most recent book, may be more approachable for some web designers, and [13] is also available. It is hoped that more books in this important subject area will be produced in the future.

## CONCLUSIONS

This paper is intended to help raise awareness of the issues concerning disabled access online, especially in the context of cultural heritage organizations such as museums that are increasingly developing their online resources with ever-more sophisticated web technologies.

For people with learning disabilities, visual disabilities, and reading impairments, print-based text can be completely inaccessible. While in recent years software developers have created electronic screen readers that convert text to speech, few of these programs offer effective control over how the text is displayed and read, nor do they provide flexible reading features. Therefore, for those with visual impairments, learning disabilities, reading disabilities, or language proficiency problems, even electronic text can be difficult to decipher. The World Wide Web poses additional barriers; while the web provides a great deal of useful, educational information, its reading levels, page design, and emphasis on graphics can make it inaccessible or unusable for some.

There is much room for improvement in the reading and speaking qualities of screen readers. The museum website accessibility survey mentioned here [9] has shown that in order to develop better accessibility the emphasis must be on improved web page coding practice. The coding aspects of web pages is extremely important in ensuring wide accessibility of websites that will be useful to all, both able bodied and disabled alike. This is possible with care and thought, but most web design professionals have yet to attain the skills to do this. It is hoped that this paper will at least raise some awareness and interest in the issues involved, particularly for cultural institutions and other public bodies that pride themselves in their physical accessibility.

The Museophile initiative [www.museophile.com], a spinout from London South Bank University, aims to help museums online in areas such as accessibility, discussion forums and collaborative e-commerce. In particular, for online information on web accessibility relevant to museums, see:

http://access.museophile.net

## ACKNOWLEDGEMENTS

Giuseppe Micheloni undertook the survey reported here as a final year project at London South Bank University [8].

Table 1. Bobby validation results for museum websites.

| Bobby | UK | | International | |
|---|---|---|---|---|
| Priority 1 | | | | |
| No errors | 10 | 40 | 7 | 28 |
| 1 error | 12 | 48 | 16 | 64 |
| 2–5 errors | 1 | 4 | 2 | 8 |
| Triggered items | 20 | 80 | 19 | 76 |
| Non-triggered | 25 | 100 | 25 | 100 |
| Priority 2 | | | | |
| No errors | 1 | 4 | 5 | 20 |
| 1 error | 5 | 20 | 5 | 20 |
| 2–5 errors | 19 | 76 | 17 | 68 |
| Triggered items | 25 | 100 | 24 | 96 |
| Non-triggered | 25 | 100 | 25 | 100 |
| Priority 3 | | | | |
| No errors | 2 | 8 | 2 | 8 |
| 1 error | 5 | 20 | 8 | 32 |
| 2–5 errors | 17 | 68 | 16 | 64 |
| Triggered items | 25 | 100 | 25 | 100 |
| Non-triggered | 25 | 100 | 25 | 100 |